\documentstyle[12pt,epsf]{article}
\newcommand{\tdm}[1]{\mbox{\boldmath $#1$}}

\newcommand{\gev}{\rm \; GeV}

\begin{document}
%
\begin{titlepage}
\pagestyle{empty}
\vspace*{1cm}
\begin{center}
{\large\bf
The QCD pomeron in $e^+ e^-$ collisions\footnote{
          Presented by L. Motyka at the Cracow Epiphany Conference on
          Electron--Positron Colliders, Cracow 5--10 Jan. 1999. }
}
\vspace{1.1cm}\\
         {\sc J.~Kwieci\'nski}$^a$,
         {\sc L.~Motyka}$^b$,
\vspace{0.3cm}\\
$^a${\it Department of Theoretical Physics, \\
H.~Niewodnicza\'nski Institute of Nuclear Physics,
Cracow, Poland}
\vspace{0.3cm}\\
$^b${\it Institute of Physics, Jagellonian University,
Cracow, Poland}
\end{center}
\vspace{1.5cm}
\begin{abstract}
The contribution of the QCD pomeron to the processes: $e^+ e^- \to e^+ e^- J/\psi J/\psi$  
and $e^+ e^- \to e^+e^- $~hadrons (with tagged electrons) is discussed.  
We focus on reactions which occur via photon-photon collisions, with virtual
photons coming from the Weizs\"{a}cker-Wiliams spectrum of the electrons.
We stress the importance of the non-leading corrections to the BFKL equation
and take  into account dominant non-leading effects which come from the
requirement that the virtuality of the exchanged gluons along the gluon
ladder is controlled by their transverse momentum squared. 
The $\gamma^*\gamma^*$ cross-sections are found to increase with increasing 
$\gamma^*\gamma^*$ CM energy $W$ as $(W^2)^{\lambda_P}$
while the cross-section for $\gamma\gamma \to J/\psi J/\psi$ is found to
increase as $(W^2)^{2\lambda_P}$. The parameter
$\lambda_P$ is slowly varying with energy $W$ and takes the values 
$\lambda_P \sim 0.23 - 0.35$ depending on the process. We also analyze the
contribution of the soft pomeron for the total $\gamma^* \gamma^*$
cross-section. We compare results of our calculations to the recent data
from LEP. 
\end{abstract}
 
\vspace{1cm}

\noindent
{\sf TPJU--4/99} \\
{\sf April 1999}\\
\end{titlepage}

\section{Introduction}

Two photon reactions are an important part of physics which is being
studied in current $e^+e^-$ experiments at LEP1 and LEP2 and which will also 
be intensively analyzed in future $e^+e^-$ colliders.
The available photon-photon energy and photon virtualities continously
increase with the increasing energy of the $e^+e^-$ pair. Therefore the
data from LEP1 and LEP2 and the expected results from the TESLA and NLC
provide us with an excellent  oportunity to study virtual photon
scattering in the diffractive regime. Moreover, with proper
experimental cuts, it is possible to study observables dominated by the
perturbative QCD contributions. The theoretical description of such
processes is based on expectations concerning high energy limit in perturbative QCD 
which is at present theoretically fairly well understood \cite{GLR,LIPAT1}. The
leading high energy behaviour is controlled by the pomeron singularity which
corresponds to the sum of ladder diagrams with reggeized gluons along
the chain. This sum is described by the Balitzkij, Fadin, Kuraev,
Lipatov (BFKL) equation \cite{BFKL1}.

The perturbative QCD pomeron exchange effects can be observed only in
specific conditions and even then not in the unambigous form.
In order to minimize the contribution of the other mechanisms competing with
the QCD pomeron and to guarantee the validity of the calculations based on
perturbative QCD one has to chose carefully the processes to analyze.
The virtualities of the gluons along the
ladder should be large enough to assure the applicability of the
perturbative expansion. The neccesary hard scale may be provided either
by coupling of the ladder to scattering particles, that contain a hard
scale themselves, or by large momentum transfer carried by the gluons.
Moreover, to distinguish the genuine BFKL from DGLAP evolution effects
it is convenient to focus on procesess in which the scales on both
ends of the ladder are of comparable size. Finally, one requires that the
non-perturbative effects should factor out in order to minimize the theoretical
uncertainties.

The two classical processes which can probe the
QCD pomeron in $ep$ and in $\gamma^* p$ collisions are the deep inelastic events
accompanied by an energetic (forward) jet {\cite{MUELLERJ, DISJET}
and the production of large $p_T$ jets separated by the rapidity gap
\cite{JETGAP}. The former process probes the QCD pomeron in the
forward direction while the latter reflects the elastic scattering of
partons via the QCD pomeron exchange with non-zero (and large)
momentum transfer. Another possible probe of the QCD pomeron at
(large) momentum transfers can be provided by the diffractive vector
meson photoproduction accompanied by proton dissociation in order to
avoid nucleon form-factor effects \cite{FORSHAW,BARTLQ}.

In this talk we shall analyze two measurements in $e^+ e^-$ collisions, 
complementary to those listed above. 
Namely we focus on double diffractive $J/\psi$ production in
$\gamma\gamma$ collisions and on the total $\gamma^* \gamma^*$ cross section.
The former process is unique since in principle it allows to test the
QCD pomeron for arbitrary momentum transfers \cite{KMPSI}.
The hard scale is given by the relatively large mass of the $c$-quark.
The total $\gamma^*\gamma^*$ cross-section has been studied by several
authors \cite{GGSTAR,BRODSKY}, however
our approach has the novel feature of taking into account dominant non-leading
corrections to the BFKL equation. This re-analysis has become necessary when
the next-to-leading corrections to the BFKL kernel were obtained
\cite{BFKLNL}, which alter substantially the results obtained at the leading
order. 
It turns out that the magnitude of the next-to-leading (NLO), 
i.e. $O(\alpha_s^2)$, 
contribution to the QCD pomeron intercept is very large for the values of the 
QCD coupling within the range which is relevant for most experiments. This 
means that the NLO approximation alone is not reliable and one has to perform 
resummation to all orders.  Unfortunately the exact result of this resummation is 
 unknown.  
It may however be possible to pin down certain dominant contributions of well 
defined physical origin and  perform their exact resummation \cite{RESUM,KMSG}.  
In our approach we shall use the so called consistency constraint which limits 
the available phase space for the real gluon emission by imposing the requirement 
that the virtuality of the exchanged gluons along the chain is dominated by their 
transverse momentum squared.  Let us remind that the form of the LO BFKL kernel 
where the gluon propagators contain only the gluon transverse momentum squared 
etc. is only valid within the region of phase space restricted by this  
 constraint.  Formally however, the consistency constraint generates 
subleading corrections.  It can be shown that at the NLO accuracy it generates 
about 70 \% of the exact result for the QCD pomeron intercept.  The very important 
merit of this constraint is also the fact that it automatically generates 
resummation of higher order contributions which stabilizes the solution
\cite{KMSG}.

\section{The total  ${\gamma^* \gamma^*}$ cross-section}

The collisions of virtual photons may be studied experimentally only
as subprocesses of reactions between charged particles. In
principle, one is able to unfold the photonic cross-section from the
leptonic data, however this procedure requires additional assumptions which
increase the systematic uncertainty of the result. It seems to be
more sensible to formulate the predictions for the $e^+e^-$ cross-sections
with the properly chosen cuts and compare them directly with the $e^+ e^-$ data.
Therefore we use the equivalent photon approximation which allows us to
express the leptonic cross-section through a convolution of the photonic
cross-section and the standard flux factors.
Thus the cross-section for the process $e^+e^- \rightarrow e^+e^- + X$
(averaged over the angle $\phi$ between the lepton scattering planes in
the frame in which the virtual photons are aligned along the $z$ axis)
is given by the following formula \cite{BRODSKY}:
$$
{Q_1^2 Q_2^2 d\sigma \over dy_1 dy_2 dQ_1^2 dQ_2^2} =
\left({\alpha\over 2 \pi}\right)^2
[P^{(T)}_{\gamma/e^+}(y_1)P^{(T)}_{\gamma/e^-}(y_2)
\sigma^{TT}_{\gamma^* \gamma^*}(Q_1^2,
Q_2^2,W^2)+
$$
$$
P^{(T)}_{\gamma/e^+}(y_1)P^{(L)}_{\gamma/e^-}(y_2)
\sigma^{TL}_{\gamma^* \gamma^*}(Q_1^2,Q_2^2,W^2)+
P^{(L)}_{\gamma/e^+}(y_1)P^{(T)}_{\gamma/e^-}(y_2)
\sigma^{LT}_{\gamma^* \gamma^*}(Q_1^2,Q_2^2,W^2)+
$$
\begin{equation}
P^{(L)}_{\gamma/e^+}(y_1)P^{(L)}_{\gamma/e^-}(y_2)
\sigma^{LL}_{\gamma^* \gamma^*}(Q_1^2, Q_2^2,W^2)]
\label{conv}
\end{equation}
where
\begin{equation}
P^{(T)}_{\gamma/e}(y) = {1 + (1-y)^2\over y}
\label{pt}
\end{equation}
\begin{equation}
P^{(L)}_{\gamma/e}(y) = 2{1-y\over y}
\label{pl}
\end{equation}
where $y_1$ and $y_2$ are the longitudinal momentum fractions of the parent
leptons carried by virtual photons,  $Q_i^2 = -q_i^2$
($i=1,2$) where $q_{1,2}$ denote the four momenta of the
virtual photons and $W^2$ is the total CM energy squared of the
two (virtual) photon system, i.e. $W^2=(q_1+q_2)^2$. The cross-sections
$\sigma^{ij}_{\gamma^* \gamma^*}(Q_1^2,Q_2^2,W^2)$ are the total
cross-sections for the process $\gamma^* \gamma^* \rightarrow X$
and the indices $i,j=T,L$ denote the polarization of
the virtual photons.  The functions $P^{(T)}_{\gamma/e}(y)$ and
$P^{(L)}_{\gamma/e}(y)$ are the transverse and longitudinal photon
flux factors.

\noindent
\begin{figure}
\leavevmode
\begin{center}
\parbox{6.5cm}{
{\large a)}\\
\epsfxsize = 5.5cm
\epsfysize = 4.2cm
\epsfbox[60 545 311 741]{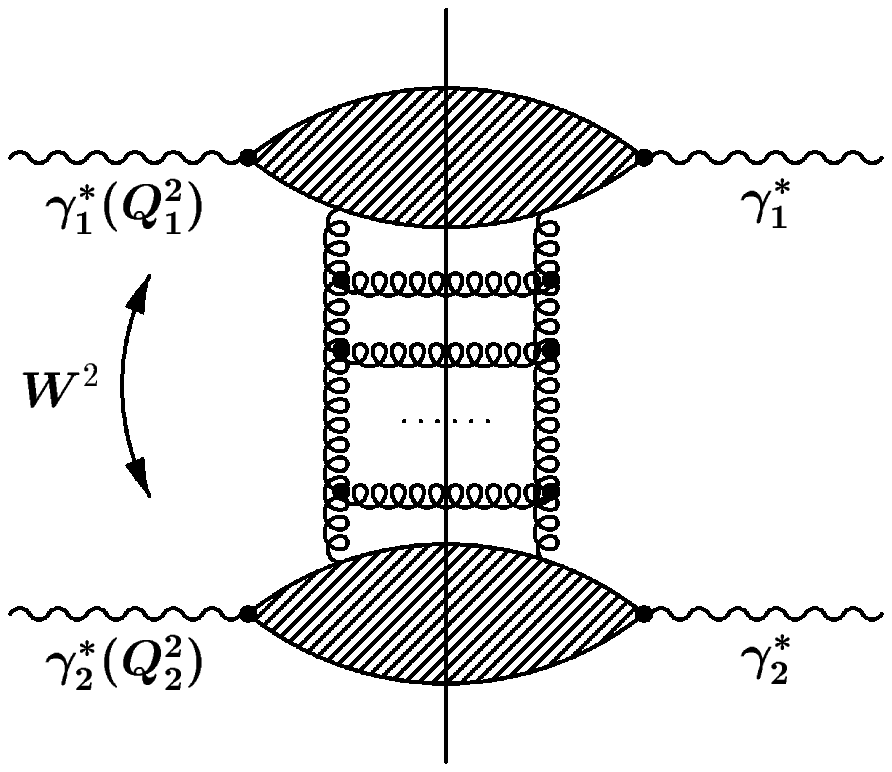}}\qquad
\parbox{6.5cm}{
{\large b)}\\
\epsfxsize = 5.5cm
\epsfysize = 4.2cm
\epsfbox{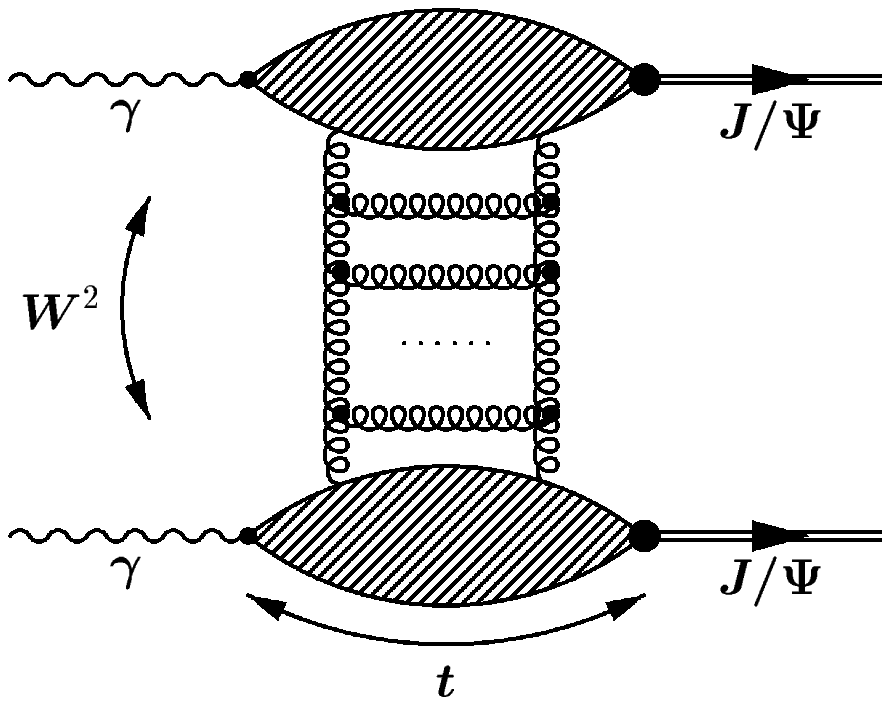}} \\
\end{center}
\caption{\small The QCD pomeron exchange mechanism of the processes
         a) $\gamma_1 ^* (Q_1 ^2) \gamma_2 ^* (Q_2 ^2) \to X$ and b)
         $\gamma\gamma \rightarrow J/\psi J/\psi$.}
\end{figure}

The ladder diagram corresponding to the perturbative contribution to
the diffractive subprocess  $\gamma_1 ^* (Q_1 ^2) \gamma^* (Q_2 ^2) \to X$
is shown in Fig.~1a.
The cross-sections  $\sigma^{ij}_{\gamma^* \gamma^*}(Q_1^2,Q_2^2,W^2)$
are given by the following formulae:
$$
\sigma^{ij}_{\gamma^* \gamma^*}(Q_1^2,
Q_2^2,W^2) = P_S(Q_1^2,Q_2^2,W^2)\delta_{iT}\delta_{jT} +
$$
\begin{equation}
{1\over 2 \pi}\sum_q\int_{k_0^2}^{k_{max}^2(Q_2^2,x)} {d^2k\over \pi k^4}
\int _{\xi_{min}(k^2,Q_2^2)}^{1/x} d\xi
 G^{0j}_q(k^2,Q_2^2,\xi)
 \Phi_i(k^2,Q_1^2,x\xi)
\label{csx}
\end{equation}
where
\begin{equation}
k_{max}^2(Q_2^2,x)=-4m_q^2+Q_2^2 \left( {1\over x} - 1 \right)
\label{kmax}
\end{equation}
\begin{equation}
\xi_{min}(k^2,Q^2)=1+{k^2+4m_q^2\over Q^2}
\label{ximin}
\end{equation}
and
\begin{equation}
x={Q_2^2\over 2q_1q_2}
\label{x}
\end{equation}
In Eq.~(\ref{csx}) we sum over four quark flavours with $m_q \to 0$ for
light quarks and $m_c=1.5\gev$.
The lower limit of integration over $k^2$ appearing in Eq.~(\ref{csx})
is taken to be $k_0 ^2 = 1\gev ^2$ in order to subtract the contribution
from the nonperturbative region from the perturbative part of the
amplitude.
The functions $G^{0i}_q(k^2,Q^2,\xi)$ are defined as
below: \cite{BRODSKY,KMSTAS}
$$
 G^{0T}_q(k^2,Q^2,\xi)=
$$
$$
2 \alpha_{em} \alpha_s(k^2+m_q^2)e_q^2\int_0^{\lambda_{max}} d \lambda
\int {d^2p^{\prime }\over \pi}\;
\delta\left[\xi-\left(1+{p^{\prime 2}+m_q^2\over z(1-z)Q^2} +
{k^2\over Q^2}\right)\right] \times
$$
\begin{equation}
\left\{ \left[(z^2 + (1-z)^2)\left({\tdm p\over D_1} -
{\tdm p + \tdm k\over D_2} \right)^2 \right]
+m_q^2 \left( {1\over D_1} - {1\over D_2} \right)^2 \right\}
\label{g0t}
\end{equation}
$$
 G^{0L}_q(k^2,Q^2,\xi)=
$$
$$
8 \alpha_{em} \alpha_s(k^2+m_q ^2)e_q^2\int_0^{\lambda_{max}} d \lambda
\int {d^2p^{\prime }\over \pi}\;
\delta\left[\xi-\left(1+{p^{\prime 2}+m_q^2\over z(1-z)Q^2} +
{k^2\over Q^2}\right)\right]  \times
$$
\begin{equation}
\left[z^2 (1-z)^2 \, \left({1\over D_1} -
{1\over D_2} \right)^2 \right]
\label{g0l}
\end{equation}
where
\begin{equation}
z={1+\lambda\over 2}
\label{zlam}
\end{equation}
\begin{equation}
\tdm p=\tdm p^{\prime} + (z-1) \tdm k
\label{pprime}
\end{equation}
$$
D_1=p^2 + z(1-z)Q^2 + m_q^2
$$
\begin{equation}
D_2=(p+k)^2 + z(1-z)Q^2 + m_q^2
\label{d12}
\end{equation}
In the formulae given above as well as throughout the rest of the
text we are using the one loop approximation for the QCD coupling
$\alpha_s$ with the number of flavours $N_f=4$ and set
$\Lambda_{QCD}=0.23 {\rm \; GeV}$.
The function $P_S(Q_1^2,Q_2^2,W^2)$  corresponds to the
contribution from the region $k^2 \le k_0^2$ in the
corresponding integrals over the gluon transverse momenta.  It
is assumed to be dominated by the soft pomeron contribution
which is estimated from the factorisation of its couplings, i.e.
\begin{equation}
P_S(Q_1^2,Q_2^2,W^2) = {\sigma^{SP}_{\gamma^*(Q_1^2)p}(Q_1^2,W^2)
\sigma^{SP}_{\gamma^*(Q_2^2)p}(Q_2^2,W^2)\over \sigma_{pp}^{SP}}
\label{sp}
\end{equation}
We assume that this term is only contributing to the transverse
part. In equation (\ref{sp}) the cross-sections
$\sigma^{SP}_{\gamma^*(Q_i^2)p}(Q_i^2,W^2)$ and
$\sigma_{pp}^{SP}$  are the soft pomeron contributions to
the $\gamma^*p$ and $pp$ total cross sections and their
parametrisation is taken from Refs. \cite{DLTOTCX,DLDIS}.  Their
$W^2$ dependence is, of course, universal i.e.
$$
\sigma_{pp}^{SP}=\beta_p^2\left({W^2\over W_0^2}\right)^{\alpha_{SP}(0)-1}
$$
\begin{equation}
\sigma^{SP}_{\gamma^*(Q_i^2)p}(Q_i^2,W^2) = \beta_{\gamma^*}(Q^2)\beta_p
\left({W^2\over W_0^2}\right)^{\alpha_{SP}(0)-1}
\label{sppar}
\end{equation}
with $W_0 = 1\gev$ and $\alpha_{SP}(0) \approx 1.08$.
The function  $  \Phi_T(k^2,Q^2,x_g) $ satisfies the Balitzkij,
Fadin, Kuraev, Lipatov  (BFKL) equation   which, in the leading
$\ln (1/x)$ approximation has the following
form:
$$
\Phi_i(k^2,Q^2,x_g)=\Phi^0_i(k^2,Q^2,x_g)+\Phi^S (k^2,Q^2,x_g)\delta_{iT}+
{3\alpha_s(k^2)\over \pi} k^2\int_{x_g}^1 {dx^{\prime}\over x^{\prime}}
\int_{k_0^2}^{\infty} {dk^{\prime 2} \over k^{\prime 2}}
$$
\begin{equation}
\left [ {\Phi_i(k^{\prime 2},Q^2,x^{\prime}) - \Phi_i(k^{ 2},Q^2,x^{\prime})
\over |k^{\prime 2} - k^{ 2}|} + {\Phi_i(k^{ 2},Q^2,x^{\prime})\over
\sqrt{4 k^{\prime 4} +  k^{4}}}\right]
\label{bfklll}
\end{equation}

In what follows we shall consider the modified BFKL equation in
which we restrict the available phase-space in the real gluon
emission by the consistency constraint:
\begin{equation}
k^{\prime 2} \le k^2{ x^{\prime}\over x_g}
\label{cc}
\end{equation}
This constraint follows from the requirement that the virtuality
of the exchanged gluons is dominated by their transverse momentum
squared.  The consistency constraint (\ref{cc}) introduces the
non-leading $\ln(1/x)$ effects and in the next-to-leading
approximation exhausts about 70\% of the entire next-to-leading corrections
to the QCD pomeron intercept.  The modiffied BFKL equation takes the
following form:
$$
\Phi_i(k^2,Q^2,x_g)=\Phi^0_i(k^2,Q^2,x_g)+\Phi^S(k^2,Q^2,x_g)\delta_{iT}+
{3\alpha_s(k^2)\over \pi} k^2\int_{x_g}^1 {dx^{\prime}\over x^{\prime}}
\int_{k_0^2}^{\infty} {dk^{\prime 2} \over k^{\prime 2}}
$$
\begin{equation}
\left [ {\Phi_i(k^{\prime 2},Q^2,x^{\prime})\Theta
\left(k^2{ x^{\prime}\over x_g}
-k^{\prime 2}\right) - \Phi_i(k^{ 2},Q^2,x^{\prime})
\over |k^{\prime 2} - k^{ 2}|} + {\Phi_i(k^{ 2},Q^2,x^{\prime})\over
\sqrt{4 k^{\prime 4} +  k^{4}}}\right]
\label{bfklcc}
\end{equation}
The inhomogeneous terms in equations (\ref{bfklll}, \ref{bfklcc})
are the sum of two contributions  $\Phi^0_i(k^2,Q^2,x_g)$
and $\Phi^S(k^2,Q^2,x_g)\delta_{iT}$.  The first term
$\Phi^0 _i (k^2, Q^2, x_g)$
corresponds to the diagram in which the two gluon system couples
to a virtual photon through a quark box and are given by
following equations:
%
%
\begin{equation}
\Phi^0_i(k^2,Q^2,x_g)=\sum_q \int_{x_g}^1 dz \; \tilde G^0_{iq}(k^2,Q^2,z)
\label{phii0}
\end{equation}
where
$$
\tilde G^0_{Tq}(k^2,Q^2,z)=2\alpha_{em} e_q^2\alpha_s(k^2+m_q^2)
\int_0^1 d\lambda \left\{
{[\lambda^2 + (1-\lambda)^2][z^2+(1-z)^2] k^2 \over \lambda(1-\lambda)k^2+
z(1-z)Q^2 + m_q^2} \right. +
$$
\begin{equation}
2m_q^2 \left. \left[ {1\over z(1-z)Q^2 + m_q^2} -
{1\over \lambda(1-\lambda)k^2+z(1-z)Q^2 + m_q^2}\right] \right\}
\label{tgt0}
\end{equation}
$$
\tilde G^0_{Lq}(k^2,Q^2,z)=16\alpha_{em}Q^2 k^2 e_q^2\alpha_s(k^2+m_q^2)
\times
$$
\begin{equation}
\int_0^1d\lambda
\left\{ {[\lambda (1-\lambda)][z^2(1-z)^2] \over [\lambda(1-\lambda)k^2+
z(1-z)Q^2 + m_q^2][z(1-z)Q^2 + m_q^2]} \right\}
\label{tgl0}
\end{equation}
The second term $\Phi^S(k^2,Q^2,x_g)\delta_{iT}$, which is
assumed to contribute only to the transverse component,
corresponds to the contribution to the BFKL equation from the
nonperturbative soft region $k^{\prime 2} < k_0^2$.  Adopting
the strong ordering approximation $k^{\prime 2} \ll k^2$ it
is given by the following formula:
\begin{equation}
\Phi^S (k^2,Q^2,x_g)=
{3 \alpha_s(k^2)\over \pi}
\int_{x_g}^1 {dx^{\prime}\over x^{\prime}}
\int_{0}^{k_0^2} {dk^{\prime 2} \over k^{\prime 2}}
 \Phi_T(k^{\prime 2},Q^2,x^{\prime})
\label{stophi}
\end{equation}
The last integral in equation (\ref{stophi}) can be interpreted
as a gluon distribution in a virtual photon of virtuality $Q^2$
evaluated at the  scale $k_0^2$.  At low values of $x^{\prime}$
it is assumed to be dominated by a soft pomeron contribution and
can be estimated using the factorisation of the soft pomeron
couplings:
\begin{equation}
\int_{0}^{k_0^2} {dk^{\prime 2} \over k^{\prime 2}}
 \Phi_T(k^{\prime 2},Q^2,x^{\prime})=\pi^2 x^{\prime}g_p(x^{\prime},k_0^2)
{\beta_{\gamma^*}(Q^2)\over \beta_p}
\label{inhos}
\end{equation}
where $g_p(x^{\prime},k_0^2)$ is the gluon distribution in a
proton at the scale $k_0^2$
and the couplings  $\beta_{\gamma^*}(Q^2)$ and $\beta_p$
are defined by equation (\ref{sppar}).
We adopt the parametrization of the gluon
structure function taken from Ref.\cite{KMSTAS} i.e.
$xg(x,k_0 ^2) = 1.57 (1-x)^{2.5}$ which is consinstent with the DIS data. 
\noindent
\begin{figure}[hbpt]
\begin{center}
\epsfxsize = 13cm
\epsfysize = 13cm
\epsfbox[18 260 555 775]{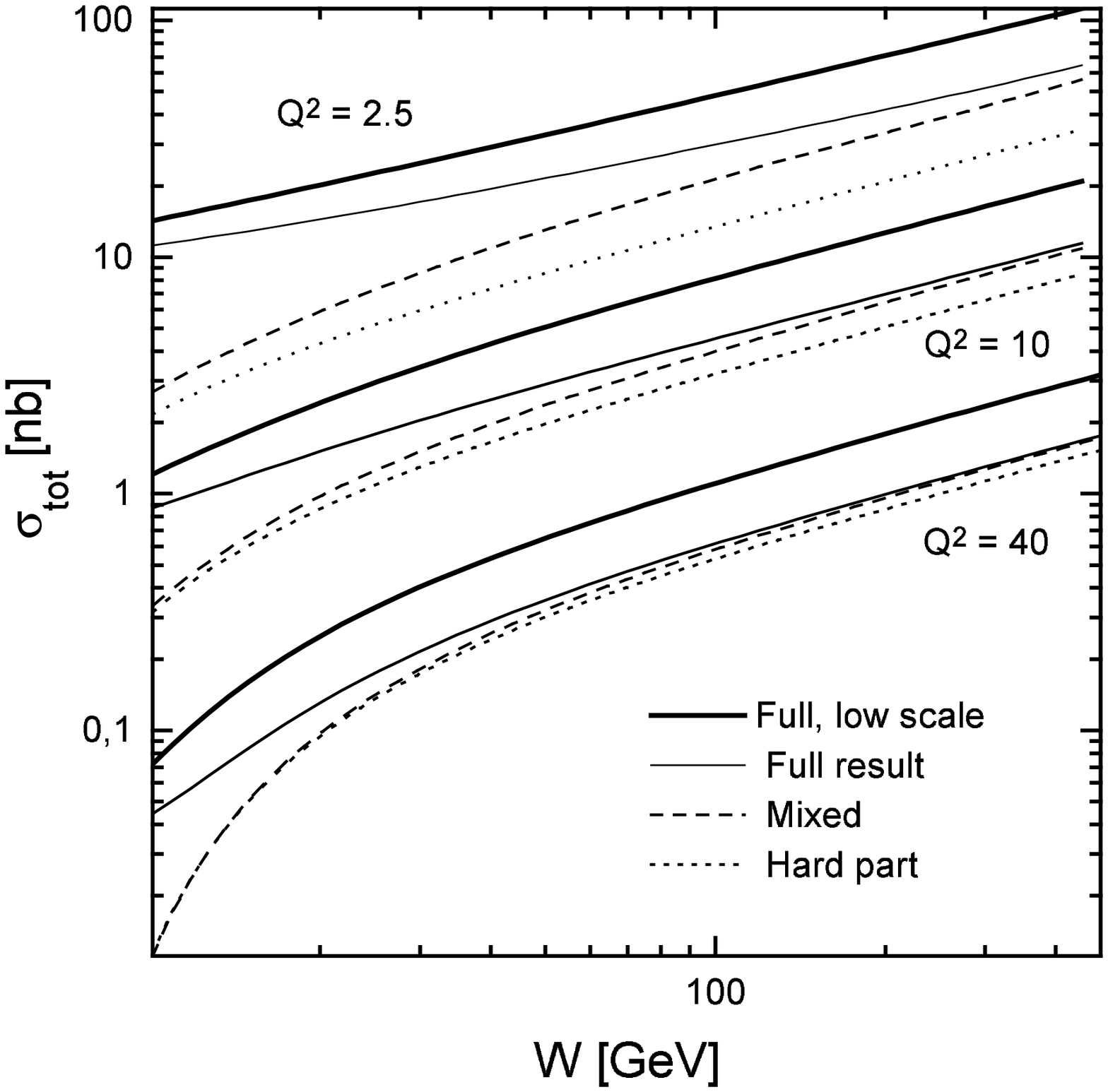}
\end{center}
\caption{\small
Energy dependence of the cross-section
$\sigma^{TT} _{\gamma^* \gamma^*}(Q_1 ^2, Q_2 ^2, W^2)$
for the process $\gamma^*(Q_1 ^2) \gamma^*(Q_2^2)  \rightarrow X$
for various choices of virtualities
$Q^2 = Q_1 ^2 = Q_2 ^2$ corresponding to Eq.~(\ref{csx}).
For each choice of the virtuality 
four curves are shown taking into account hard effects only (``hard part''),
hard amplitude with soft pomeron contributions added in the source term
of the BFKL equation (``mixed''),
the full cross-section including both soft and hard pomeron
contributions (``full result''). We also show the ``full result'' with the
low scale of $\alpha_s$ in the impact factors:
$\mu^2 = (k^2 + m_q^2)/4$.
}
\end{figure}

In Fig.~2 we show our results for 
$\sigma^{TT}_{\gamma^* \gamma^*} (Q_1 ^2, Q_2 ^2, W^2)$ plotted
as the function of the CM energy $W$ for three different values of  $Q^2$
where $Q_1^2 = Q_2^2 = Q^2$.  We plot in this figure:
\begin{enumerate}
\item the pure QCD
(i.e. ``hard'') contribution obtained from solving the BFKL equation with the
consistency constraint included (see Eq.~(\ref{bfklcc})) and with the
inhomogeneous term containing only the QCD impact factor defined by equations
(\ref{phii0},\ref{tgt0},\ref{tgl0}),
\item
the ``mixed" contribution generated by the BFKL equation (\ref{bfklcc})
with the soft pomeron contribution defined by equations (\ref{stophi},
\ref{inhos}) included in the inhomogeneous term,
\item
The ``full" contribution which also contains the soft pomeron term (\ref{sp}).
\end{enumerate}
We also show results obtained by changing the scale of the strong coupling
$\alpha_s$ in the impact factors from $k^2+m_q^2$ to $(k^2+m_q^2)/4$.
The scale of $\alpha_s$ in the BFKL equation is the same in the both cases.
The components of the cross-section for which at least one of the photons
is longitudinally polarized have very similar energy dependence to
$\sigma^{TT} _{\gamma^* \gamma^*} (Q_1^2,Q^2_2,W^2)$ 
and give together about 60\% of the transverse-transverse contribution.

We see from this figure that the effects of the soft pomeron contribution
are non-negligible at low and moderately large values of $Q^2 < 10 \gev^2$
and for moderately large values of $W < 100 \gev$.
The QCD pomeron however dominates already at $Q^2=40 \gev^2$.
We also see from this figure that for low energies $W<40 \gev$ the
phase-space effects are very important.
For $W>40\gev$ or so one observes that the cross-section exhibits
the effective power-law behaviour
$\sigma_{\gamma^* \gamma^*}(W) \sim (W^2)^{\lambda_P}$.
The (effective) exponent increases weakly
with increasing $Q^2$ and varies from $\lambda_P=0.28$ for $Q^2=2.5\gev^2$ to
$\lambda_P=0.33$ for $Q^2 = 40 \gev^2$.  This (weak) dependence of the
effective exponent $\lambda_P$ with $Q^2$ is the result of the interplay
between soft and hard pomeron contributions,
where the former becomes less important at large $Q^2$.\\

\begin{table}
\caption{ Comparison of the theoretical results to L3 data for $e^+ e^- \to e^+ e^- X$ 
with  $E_{tag} > 30$ GeV, 30 mrad $ < \theta_{tag}  <$ 66 mrad. We show in
the table $d\sigma / dY$ binned in $Y$ obtained from experiment and 
the results of our calculation which take into account perturbative pomeron only
(hard) and both perturbative and soft pomerons (hard + DL) for two different
choices of scale of the $\alpha_s$ in impact factors and for $e^+e^-$ CM
energy 91~GeV and 183~GeV.}

\begin{center}
\begin{tabular}{||c|c||c|c||c|c||}
\hline\hline 
 &  \multicolumn{5}{|c||}{$\langle d\sigma / dY \rangle$ [fb] }\\
\cline{2-6} 
  & & \multicolumn{4}{c||}{ Theory (BFKL+DL)} \\
 \cline{3-6}  
$\Delta Y$ & Data --- QPM & \multicolumn{2}{c||}{$\alpha_s[(\tdm k^2+m_q^2)/4]$} &
       \multicolumn{2}{c||}{$\alpha_s (\tdm k^2+m_q^2)$} \\
  \cline{3-6}  
&    & Hard & Hard + DL & Hard & Hard + DL \\ 
\hline\hline
\multicolumn{6}{||c||}{91 GeV} \\ \hline
2 -- 3 & $480 \pm 140 \pm 110$  & 76 & 206 & 34 & 163\\  \hline 
3 -- 4 & $240 \pm 60 \pm 50  $  & 114& 237 & 53 & 173\\ \hline
4 -- 6 & $110 \pm 30 \pm 10  $  & 60 & 109 & 29 & 74\\ \hline\hline
\multicolumn{6}{||c||}{183 GeV} \\ \hline
2 -- 3 & $180 \pm 120 \pm 50$  &  51 & 68  & 25 & 42\\  \hline 
3 -- 4 & $160 \pm 50 \pm 30  $  & 70 & 86  & 34 & 49\\ \hline
4 -- 6 & $120 \pm 40 \pm 20  $  & 70 & 85  & 35 & 47\\ \hline\hline
\end{tabular}
\end{center}
\end{table}

Using Formula~(\ref{conv}) integrated over the virtualities in the range
allowed by the relevant experimental cuts,
we have calculated the total cross-section for the process 
$e^+ e^- \rightarrow e^+ e^- + X$  for LEP1
and LEP2 energies and confronted results of our calculation with the
recent experimental data obtained by the L3 collaboration at LEP \cite{L3}.
Comparison of our results with experimental data is sumarised in Table~1.
We show comparison for  $d\sigma/dY$, where $Y=\ln (W^2/Q_1Q_2)$
with subtracted Quark Parton Model (QPM)  contribution.
We see that the contamination of the cross-section by soft pomeron
is substantial. The data do also favour the smaller value of the scale of
$\alpha_s$. In general, the results of our calculation lay below the data,
however the error bars are still quite large, so that the discrepancy
is not very pronounced. Let us also mention that cuts applied to obtain
the data shown in Table~1 admit rather low $\gamma\gamma$ energies i.e.
below 10~GeV \cite{L3}, which probably is not sufficient to justify
the validity of high energy limit in QCD.


\section{Exclusive $J/\psi$ production}

The experimental aspects of the measurement of double exclusive
$J/\psi$ production are different from those for the virtual photons
scattering. Namely, since the $c$-quark provides
the energy scale, we may perturbatively describe the cross-section for the
process of exclusive $J/\psi$ production in which almost real photons
take part. It is an important feature beacause the photon flux in electron is
dominated by low virtualities. On the other hand one may measure the
produced $J/\psi$-s through theirs decay products with no need of
tagging of the electrons.
Thus, it is prefered to focus on events with anti-tagged leptons.
The cross-section for the process $e^+e^- \rightarrow e^+e^- + Y$
for anti-tagged $e^{\pm}$ corresponds to the production of the hadronic
state~$Y$ in $\gamma\gamma$ collision and is given by the following
convolution integral: \cite{GGREV}
\begin{equation}
\sigma_{e^+e^- \to e^+e^- + Y} =
\int_0^1 dy_1 \int_0^1 dy_2 \Theta ( W^2 -  W^2_{Y0})
\sigma_{\gamma \gamma \rightarrow Y}(W^2)
f_{\gamma/e}(y_1) f_{\gamma/e}(y_2).
\label{conv2}
\end{equation}
where the $\gamma\gamma$ system invariant mass squared $W^2$ is related
to the lepton CM energy squared $s$ by the simple formula: $W^2 = y_1y_2 s$.
The flux factor takes the form:
\begin{equation}
f_{\gamma/e}(y)={\alpha_{em}\over 2 \pi}
\left[ {1 + (1-y)^2 \over y}\;{\rm ln}\,{Q_{max}^2\over Q_{min}^2} -
2 m_e ^2 y \left({1\over Q_{min}^2} - {1\over Q_{max} ^2}\right) \right].
\label{flux}
\end{equation}
and
\begin{equation}
 Q_{min}^2 =  {m_e^2 y^2 \over ( 1-y)}
\label{pmin}
\end{equation}
\begin{equation}
Q_{max}^2=(1-y)E_{beam}^2 \theta_{max}^2.
\label{pmax}
\end{equation}
The lower limit follows from the kinematics of photon emission from a lepton
whereas the upper one
arises from the upper limit $\theta_{max}$ for the lepton scattering angle.
The minimal invariant mass squared of the hadronic system $W^2_{Y0}$, the
angle $\theta_{max}$ and the beam energy $E_{beam}$ depend on the process
and experimental conditions. For diffractive $J/\psi$ production we shall
choose $\theta_{max}=30$~mrad in accordance with LEP conditions and
$W_{Y0}=15\gev$.\\

The formalism that we shall employ to evaluate the cross-section of the
sub-process $\gamma \gamma \rightarrow J/\psi  J/\psi$
is very similar to this used in the previous section.
However some modification are neccessary in order to adopt to
specific features of the process.
First of all we have to go beyond the forward configuration of the pomeron
by the use of the BFKL equation with non-zero momentum transver. Besides
that, we introduce a parameter $s_0$ in the propagators of exchanged gluons
instead of the infra-red cut-off $k_0 ^2$ applied in the previous case.
This parameter can be viewed upon as the effective representation of the
inverse of the colour confinement radius squared. Sensitivity of the
cross-section to its magnitude can serve as an estimate of the sensitivity
of the results to the contribution coming from the infrared region. It
should be noted that formula (\ref{ima}) gives finite result in the limit
$s_0=0$. While analyzing this process we use the asymptotic (high-energy)
form of the amplitude, neglecting the phase space effects.\\

The imaginary part ${\rm Im} A(W^2,t=-Q_P ^2)$ of the amplitude for
the considered process which corresponds to the diagram in Fig.~1b
can be written in the following form:
\begin{equation}
{\rm Im} A(W^2,t=- Q_P^2) =
\int {d^2\tdm k\over\pi}{\Phi_0(k^2, Q_P ^2)\Phi(x,\tdm k,\tdm Q_P)\over
[(\tdm k + \tdm Q_P /2)^2 +s_0][(\tdm k - \tdm Q_P /2)^2+s_0]}
\label{ima}
\end{equation}
In this equation $x=m_{J/\psi}^2/W^2$ where $W$ denotes the total
CM energy of the $\gamma \gamma$ system, $m_{J/\psi}$ is the
mass of the $J/\psi$ meson, $\tdm Q_P / 2 \pm \tdm k$ denote the
transverse momenta of the exchanged gluons and $\tdm Q_P$ is the
transverse part of the momentum transfer.

\noindent
\begin{figure}
\epsfxsize = 12cm
\epsfysize = 8cm
\epsfbox{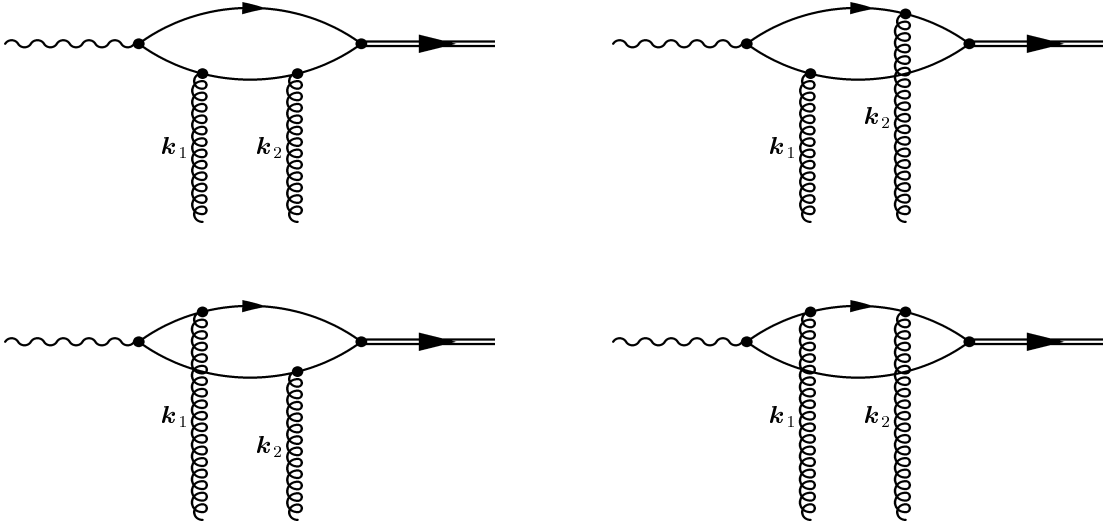}
\caption{\small
The diagrams describing the coupling of two gluons
to the $\gamma \rightarrow J/\psi$ transition vertex.
}
\end{figure}

The impact factor $\Phi_0(k^2, Q_P^2)$ describes the $\gamma J/\psi$
transition induced by two gluons and the diagrams defining
this factor are illustrated in Fig.~3. In the nonrelativistic approximation
they give the following formula for $\Phi_0(k^2, Q_P^2)$
\cite{FORSHAW,GINZBURG}:
\begin{equation}
\Phi_0(k^2, Q_P^2)=
{C\over 2}\sqrt{\alpha_{em}}\alpha_s(\mu^2) \left[{1\over \bar q^2} -
{1\over m_{J/\psi}^2/4+k^2}\right]
\label{impf0}
\end{equation}
where
\begin{equation}
C=q_c{8\over 3} \pi m_{J/\psi} f_{J/\psi}
\label{c}
\end{equation}
with $q_c=2/3$ denoting the charge of a charm quark
and
\begin{equation}
\bar q ^2= {m_{J/\psi}^2+Q_P^2\over 4}
\label{qbar2}
\end{equation}
\begin{equation}
f_{J/\psi}= \sqrt{
{3m_{J/\psi}\Gamma_{J/\psi\rightarrow l^+ l^-}
\over 2\pi \alpha_{em} ^2}
}
\label{fpsi}
\end{equation}
where $\Gamma_{J/\psi \rightarrow l^+ l^-}$ is the leptonic with of the
$J/\psi$ meson.  In our calculations we will set
$f_{J/\psi}=0.38{\rm \; GeV}$.
The function $\Phi(x,\tdm k,\tdm Q_P)$ satisfies the non-forward BFKL equation
which in the leading $\ln(1/x)$ approximation has the following form:
$$
\Phi(x,\tdm k,\tdm Q_P)=\Phi_0(k^2, Q_P^2)+ {3\alpha_s(\mu^2)\over
2\pi^2}\int_x^1{dx^{\prime}\over x^{\prime}} \int
{d^2\tdm k' \over (\tdm k' - \tdm k)^2 + s_0} \times
$$
$$
\left\{\left[{{\tdm k_1^2}\over {\tdm k_1^{\prime 2}} + s_0}   +
{{\tdm k_2^2}\over {\tdm k_2^{\prime 2}} + s_0}
  - Q_P^2
 {(\tdm k' - \tdm k)^2+s_0 \over ({\tdm k_1^{\prime 2}} + s_0)
 ({\tdm k_2^{\prime 2}} + s_0)}
\right]
\Phi(x',\tdm k' ,\tdm Q_P) - \right.
$$
\begin{equation}
\left. \left[{{\tdm k_1^2}\over {\tdm k_1^{\prime 2}}  +
(\tdm k' - \tdm k)^2 +2s_0} +
{{\tdm k_2^2}\over {\tdm k_2^{\prime 2}}  +
(\tdm k' - \tdm k)^2 +2s_0} \right]
\Phi(x',\tdm k,\tdm Q_P) \right\}
\label{bfkl}
\end{equation}
where
$$
{\tdm k_{1,2}} = {\tdm Q_P \over 2}\pm \tdm k
$$
and
\begin{equation}
{\tdm k_{1,2}^{\prime}} = {\tdm Q_P \over 2} \pm \tdm k^{\prime}
\label{k12}
\end{equation}
denote the transverse momenta of the gluons.
The scale of the QCD coupling $\alpha_s$ which appears in
equations (\ref{impf0}) and (\ref{bfkl}) will be  set
$\mu^2=k^2+Q_P^2/4 +m_c^2$ where $m_c$ denotes the mass of the
charmed quark.  The differential cross-section is related in the following
way to the amplitude~$A$:
\begin{equation}
{d \sigma \over dt} = {1\over 16 \pi} |A(W^2,t)|^2
\label{dsdt}
\end{equation}
Generalization of the consistency constraint (\ref{cc})
to the case of non-forward configuration
with $Q_P^2 \ge 0$ takes the following form:
\begin{equation}
k'^2 \le (k^2+ Q_P^2/4) {x'\over x}
\label{kc2}
\end{equation}
Besides the BFKL equation (\ref{bfkl}) in the leading
logarithmic approximation we shall also consider the
equation which will embody the constraint (\ref{kc2}) in order to
estimate the effect of the non-leading contributions. \\

The corresponding equation which contains constraint (\ref{kc2}) in the
real emission term reads:
$$
\Phi(x,\tdm k,\tdm Q_P)=\Phi_0(k^2, Q_P^2)+ {3\alpha_s(\mu^2)\over
2\pi^2}\int_x^1{dx^{\prime}\over x^{\prime}}
\int {d^2\tdm k' \over (\tdm k' - \tdm k)^2 + s_0} \times
$$
$$
\left\{\left[{{\tdm k_1^2}\over {\tdm k_1^{\prime 2}} + s_0}   +
{{\tdm k_2^2}\over {\tdm k_2^{\prime 2}} + s_0}
  - Q_P^2
 {(\tdm k' - \tdm k)^2+s_0 \over ({\tdm k_1^{\prime 2}} + s_0)
 ({\tdm k_2^{\prime 2}} + s_0)}
\right]
 \Theta \left((k^2+Q_P^2/4)x'/x-k^{\prime 2}) \right)  \times \right.
$$
\begin{equation}
\left.
\Phi(x',\tdm k',\tdm Q_P) -
\left[{{\tdm k_1^2}\over {\tdm k_1^{\prime 2}}  +
(\tdm k' - \tdm k)^2 +2s_0} +
{{\tdm k_2^2}\over {\tdm k_2^{\prime 2}}  +
(\tdm k' - \tdm k)^2 +2s_0} \right]
\Phi(x',\tdm k,\tdm Q_P) \right\}
\label{bfklkc}
\end{equation}
We solved  equations (\ref{bfkl}) and (\ref{bfklkc}) numerically setting
$m_c=m_{J/\psi} /2$.
Brief summary of the numerical method and of the adopted
approximations in solving
equations (\ref{bfkl},\ref{bfklkc}) has been given in Ref.\cite{KMPSI}.
Let us recall that we used running coupling with the scale
$\mu^2=k^2+Q_P^2/4+m_c^2$. The parameter $s_0$ was varied within the range
$0.04 {\rm \; GeV}^2 < s_0<0.16 {\rm \; GeV}^2$.  It should be noted
that the solutions of
equations (\ref{bfkl}, \ref{bfklkc}) and the amplitude (\ref{ima}) are
finite in the limit $s_0=0$. This follows from the fact that both impact
factors $\Phi_0(k^2, Q_P^2)$ and $\Phi(x,\tdm k,\tdm Q_P)$ vanish for
$\tdm k=\pm \tdm Q_P/2$ (see equations (\ref{impf0}, \ref{bfkl},
\ref{bfklkc})). The results with finite $s_0$ are however  more realistic.

\noindent
\begin{figure}[hbpt]
\leavevmode
\begin{center}
\epsfxsize = 13cm
\epsfysize = 13cm
\epsfbox[18 200 565 755]{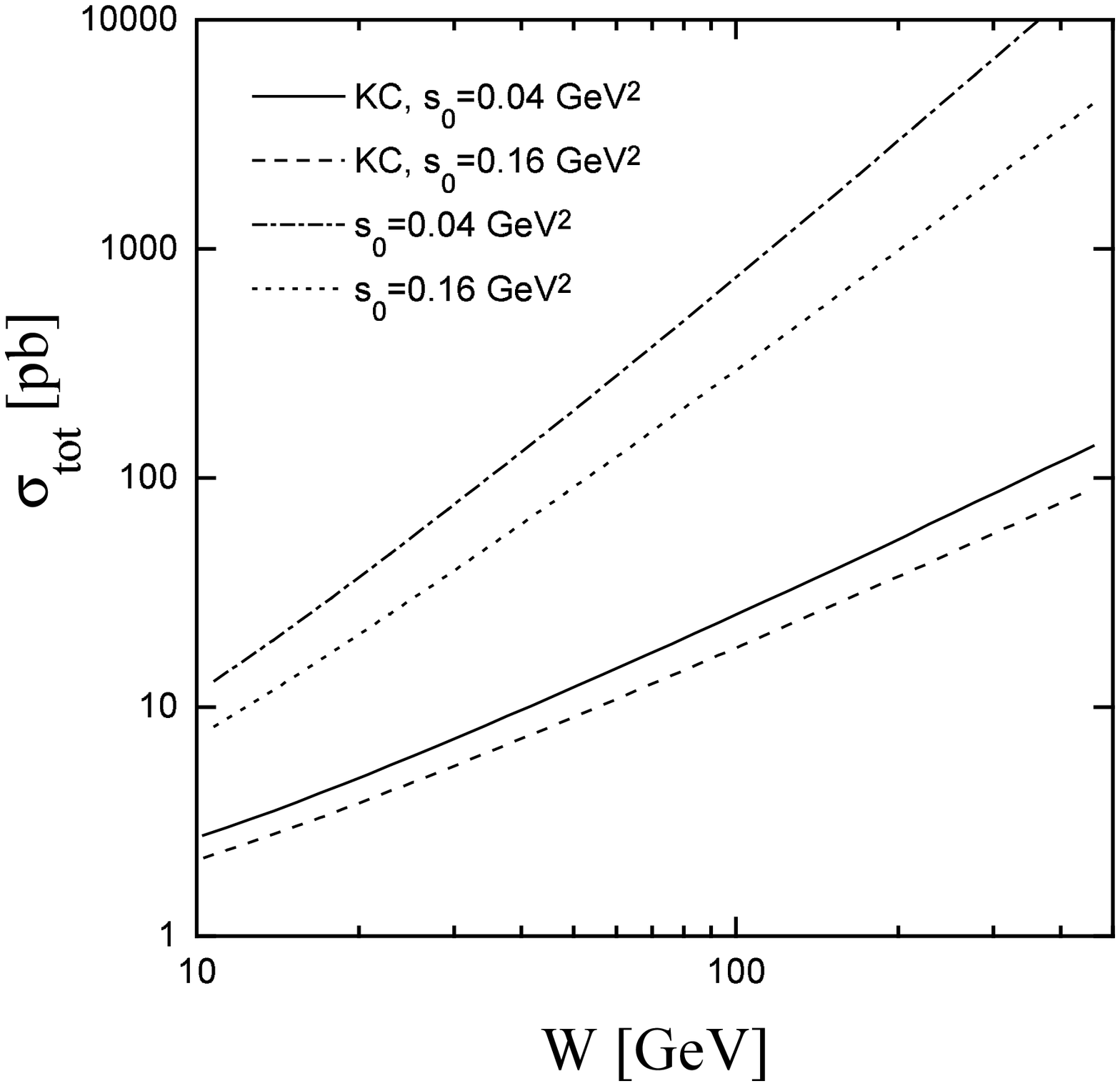}
\end{center}
\caption{\small
Energy dependence of the cross-section for the process
$\gamma\gamma \rightarrow J/\psi J/\psi$.  The two lower curves
correspond to the calculations based on  equation
(\ref{bfklkc}) which contains the non-leading effects coming
from the  constraint (\ref{kc2}).  The continuous line
corresponds to $s_0=0.04 {\rm \; GeV}^2$ and the dashed line to $s_0=0.16
{\rm \; GeV}^2$.  The two upper curves correspond to equation
(\ref{bfkl}) i.e.
to the BFKL equation in the leading logarithmic approximation.
The dashed-dotted line corresponds to $s_0=0.04 {\rm \; GeV}^2$
and short dashed line to $s_0=0.16{\rm \; GeV}^2$.
}
\end{figure}


In Fig.~4 we show the cross-section for the process
$\gamma \gamma \rightarrow J/\psi J/\psi$
plotted as the function of the total CM energy $W$.
We show results based on the BFKL equation in the leading logarithmic
approximation as well as those which include the dominant non-leading
effects.  The calculations were performed for the two values of the parameter
$s_0$ i.e. $s_0=0.04 {\rm \; GeV}^2$ and $s_0=0.16 {\rm \; GeV}^2$.
In Fig.~5 we show the $t$-dependence of the cross-section calculated for
$s_0 = 0.10 {\rm \; GeV}^2$.
We show in this figure results for two values of the CM energy $W$
($W=50 {\rm \; GeV}$  and $W=125 {\rm \; GeV}$)
obtained from the solution of the BFKL equation with the non-leading effects
taken into account (see Eq.~(\ref{bfklkc})) and confront them with
the Born term which corresponds to the two (elementary) gluon exchange.
The latter is of course independent of the energy $W$.
The values of the energy $W$ were chosen to be in the region which may
be accessible at LEP2.
Let us discuss crucial features of the obtained results:
\begin{enumerate}
\item {\bf Non leading corrections}. We see from Fig.~4 that the effect of
the non-leading contributions is very important and that they  significantly
reduce magnitude of the cross-section and slow down its increase with
increasing CM energy $W$.

\item
{\bf Energy dependence}.
The cross-section exhibits approximate $(W^2)^{2\lambda_P}$
dependence.  The  parameter $\lambda_P$, which slowly varies with
the energy~$W$ takes the values $\lambda_P \sim 0.23 - 0.28$  within the
energy range $20{\rm \; GeV} < W < 500{\rm \; GeV}$  relevant for LEP2
and for possible TESLA measurements. These results correspond to the
solution of the BFKL equation (\ref{bfklkc}) which contains the non-leading
effects generated by the constraint (\ref{kc2}).
The (predicted) energy dependence of the cross-section
($(W^2)^{2\lambda_P}, \lambda_P \sim 0.23 - 0.28$)
is  marginally steeper than  that observed in
$J/\psi$ photo-production \cite{VMPHOTOP}. It should however be remebered
that the non-leading effects which we have taken into account although
being the dominant ones still do not exhaust all next-to-leading QCD
corrections to the BFKL kernel \cite{BFKLNL}. The remaining contributions
are expected to reduce the parameter $\lambda_P$ but their effect may be
expected to be less important than that generated by the constraint
(\ref{kc2}). The cross-section calculated from the BFKL equation in the
leading logarithmic approximation gives much stronger energy dependence of
the cross-section (see Fig.~4).

\item
{\bf The value of the cross-section}.
Enhancement of the cross-section is still appreciable  after
including the dominant non-leading contribution which follows from the
constraint (\ref{kc2}). Thus while in the Born approximation
(i.e. for the elementary two gluon exchange
which gives energy independent cross-section)
we get $\sigma_{tot} \sim 1.9-2.6$~pb the cross-section calculated from the
solution of the BFKL equation with the non-leading effects taken into account
can reach the value 4~pb at $W=20 \gev$ and 26~pb for $W=100 \gev$ i.e. for
energies which can be accessible at LEP2.

\item
{\bf Infrared sensitivity}.
The magnitude of the cross-section decreases with increasing
magnitude of the  parameter $s_0$ which controls the
contribution coming from the infrared region.  This effect is
however much weaker than that generated by the
constraint (\ref{kc2}) which gives the dominant non-leading
contribution.  The  energy dependence of the
cross-section is  practically unaffected  by the parameter $s_0$.

\item
{\bf The $t$-dependence}.
Plots shown in Fig.~5 show that the BFKL effects significantly affect
the $t$-dependence of the differential cross-section leading to steeper
$t$-dependence than that generated by the Born term.   Possible
energy dependence of the diffractive slope is found to be very
weak (see Fig.~5).  Similar result was also found in the BFKL equation
in the leading logarithmic approximation \cite{BARTLQ}.
\end{enumerate}

\begin{figure}
\leavevmode
\begin{center}
\epsfxsize = 13cm
\epsfysize = 13cm
\epsfbox{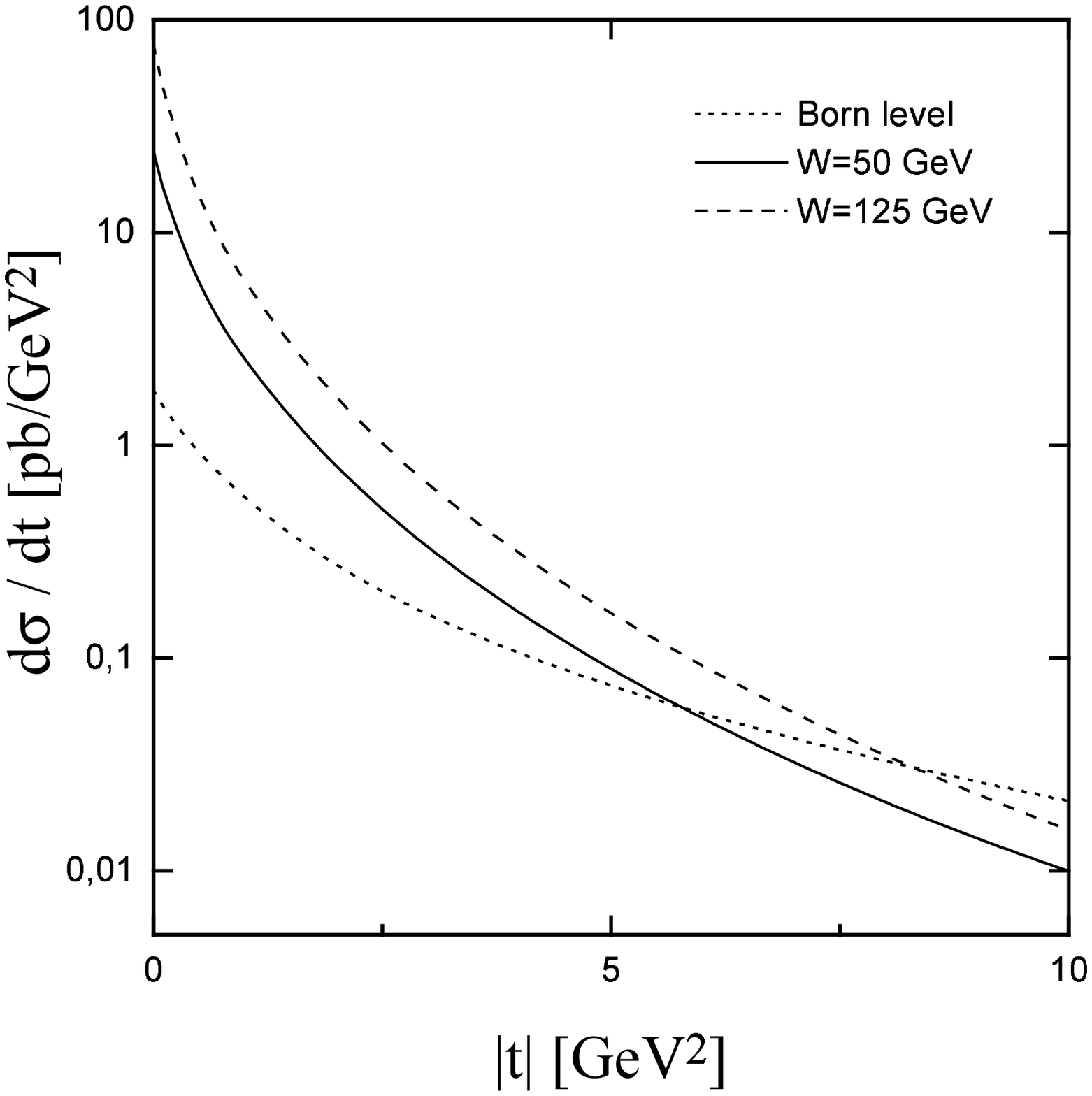}
\end{center}
\caption{\small
The differential cross-section of the process $\gamma
\gamma \rightarrow J/\psi J/\psi$ corresponding to the solution
of  equation (\ref{bfklkc}) which contains the non-leading effects coming
from the consistency (kinematical) constraint (\ref{kc2}) shown for two
values of the CM energy $W$, $W=50{\rm \; GeV}$ (continuous line) and
$W=125 {\rm \; GeV}$ (dashed line).  The short dashed line corresponds
to the Born term i.e. to the elementary two gluon exchange
mechanism which gives the energy independent cross-section.
The parameter $s_0$ was set equal to $0.10{\rm \; GeV}^2$.
}
\end{figure}

In our calculations we have assumed dominance of the imaginary part of the
production amplitude. The effect of the real part can be taken into account by
multiplying the cross-section by the correction factor
$1+tg^2(\pi\lambda_P/2)$ which for $\lambda_P \sim 0.25$ can introduce
additional enhancement of about 20~\%.\\

The photonic cross-sections that we obtained in this section are rather
low in terms of the expected number of events, at least for the LEP2
luminosity. Therefore we consider the most inclusive observables relevant for
double $J/\psi$ production in $e^+e^-$ collisions which is the total
cross-section $\sigma_{tot} (e^+ e^- \to e^+ e^- J/\psi J/\psi)$. In fact,
it is convenient to impose additionally the anti-tagging condition.
Taking $\theta_{max}=30$~mrad we get for the
$\sigma_{tot} (e^+ e^- \to e^+ e^- J/\psi J/\psi)$ the values of
about 0.14~pb at $\sqrt{s}=175\gev$ and 0.74~pb at $\sqrt{s}=500\gev$
(i.e. for typical energies at LEP2 and TESLA respectively). Therefore,
assuming the LEP2 luminosity to be about 500~pb$^{-1}$ we predict about
70~events, which is far below the previous expectations~\cite{GGREV}.
Besides, if one measures both the $J/\psi$-s through the leptonic decay
channels the rate should be divided by factor of about 20, which cuts down
the statistics to only a few events.

\section{Discussion and summary}

From the theoretical point of view, there exist excellent oportunities
to study the exchange of the QCD pomeron in $e^+ e^-$ colliders.
The two golden-plated measurements for this purpose are exclusive $J/\psi$
production and the total $\gamma^* \gamma^*$ cross-section. Both these
processes allow to reduce substantially the contribution of unknown,
nonperturbative elements. However, the leptonic cross-sections in both
cases are well below 1~pb in LEP2 conditions, which makes the measurement
rather difficult there. Nevertheless this problem does not appear at the
future linear colliders $e^+ e^-$ for which the luminosity is expected to
be much larger than at LEP and moreover the cross-section for diffractive
processes is enhanced due to the photon flux and the pomeron effects.
The large expected statistics enables one to reach the region of large
photon virtualities (for double tagged events) where the perturbative
calculations are more reliable. \\

The important point that should be stressed once more is the existence of large
non-leading corrections to BFKL equation, which influence dramatically
the theoretical estimate of the pomeron intercept i.e. the behaviour of the
cross-sections as functions of the energy. The recently calculated magnitude
of next-to-leading contribution to the intercept (for any relevant value of
the strong coupling constant) is  comparable or even greater than the leading
term. This implies a very poor convergence of the perturbative series. Thus
one is forced to rely on a resummation scheme.
We adopt the so called consistency
constraint, which is based on the requirement that the 
virtualities of gluons exchanged along the ladder are dominated by
transverse momenta squared. 
This constraint introduces at the next-to-leading order
a correction to the pomeron intercept which exhausts about 70\% of the
exact QCD result. The main advantage of this approach is that there is a
good physical motivation behind it. Moreover it also offers an approximate
resummation scheme for the perturbative expansion of the intercept.\\

Employing this scheme we found significant reduction of the predicted value of
the intercept in comparison to the leading value.
We find that the calculated behaviour of the 
$\gamma^* \gamma^*$ total cross-section exhibits approximate power law
dependence $(W^2)^{\lambda_P}$ with $0.28<\lambda_P<0.35$. It is also found  
that the cross-section for  $\gamma\gamma \to J/\psi J/\psi$ increases 
with  increasing energy $W$ as $(W^2)^{2\lambda_P}$ with $\lambda_P$ varying
from~0.23 to~0.28. This has important consequences for the phenomenology,
since the enhancement of the cross-section although
still quite appreciable is much smaller than that which follows from
estimates based on the leading logarithmic approximation \cite{GGREV}.
The results of our calculation are in fair
agreement with the existing data for $\gamma^*\gamma^*$ cross-section from
LEP, although the theoretical calculations have a tendency to underestimate
experimental results.  
They are also much more realistic than
the predictions following from the leading order BFKL
equation, which are an order of
magnitude larger. The encouraging element is that even this very first
data with rather low statistics, are enough to show clearly the importance of
non-leading corrections. We may therefore expect that when the excellent data
from linear colliders will be available we will acquire very good opportunity 
to test our models and to understand more deeply the physics of the 
QCD pomeron. \\

\section*{Acknowledgments}
We are grateful to the Organizers for the interesting and stimulating
Conference. We thank Albert De Roeck for his interest in this work and useful
discussions. This research was partially supported
by the Polish State Committee for Scientific Research (KBN) grants
2~P03B~184~10, 2~P03B~89~13, 2~P03B~084~14 and by the
EU Fourth Framework Programme 'Training and Mobility of Researchers', Network
'Quantum Chromodynamics and the Deep Structure of Elementary Particles',
contract FMRX--CT98--0194.


\begin{thebibliography}{9999}
\bibitem{GLR}  L.N. Gribov, E.M. Levin and M.G. Ryskin,
Phys. Rep. {\bf 100} (1983) 1.
\bibitem{LIPAT1} L.N. Lipatov, Phys. Rep. {\bf 286} (1997) 131.
\bibitem{BFKL1} E.A. Kuraev, L.N.Lipatov and V.S. Fadin, Zh. Eksp. Teor. Fiz.
{\bf 72} (1977) 373 (Sov. Phys. JETP {\bf 45} (1977) 199);
Ya. Ya. Balitzkij and L.N. Lipatov, Yad. Fiz. {\bf 28} (1978) 1597 (Sov. J.
Nucl. Phys. {\bf 28} (1978) 822);
J.B. Bronzan and R.L. Sugar, Phys. Rev. {\bf D17} (1978) 585;
T. Jaroszewicz, Acta. Phys. Polon. {\bf B11}
(1980) 965; L.N. Lipatov, in "Perturbative QCD", edited
by A.H. Mueller, (World Scientific, Singapore, 1989), p. 441.
\bibitem{MUELLERJ}A.H. Mueller, J. Phys. {\bf G17}  (1991) 1443 .
\bibitem{DISJET}J. Bartels, M. Loewe and A. De~Roeck, Z. Phys. {\bf C54}
(1992) 635 ;J. Kwieci\'nski, A.D. Martin and P.J. Sutton,
Phys. Rev. {\bf D46}   (1992) 921; Phys. Lett. {\bf B287}
(1992) 254;J. Bartels et al., Phys. Lett. {\bf B384} (1996) 300;
J.~Bartels, V.~Del~Duca,
M. W\"usthoff,  Z. Phys. {\bf C76} (1997) 75.
E. Mroczko,  Proceedings of the 28th
International Conference on High Energy Physics, Warsaw, Poland, 25-31 July
1996, Z. Ajduk and A.K Wr\'oblewski (editors), World Scientific.
\bibitem{JETGAP}A.H. Mueller, W.K. Tang, Phys. Lett. {\bf B284} (1992) 123;
V. Del Duca, W.K. Tang, Phys. Lett. {\bf B312} (1993) 225; V. Del Duca,
C.R. Schmidt, Phys.Rev. {\bf D49} (1994) 4510.
\bibitem{FORSHAW}J.R. Forshaw, M.G. Ryskin, Z. Phys. {\bf C68} (1995) 137.
\bibitem{BARTLQ} J. Bartels,  J.R. Forshaw , H. Lotter, M. W\"usthoff,
 Phys. Lett. {\bf B375} (1996) 301.
%
\bibitem{KMPSI} J.~Kwieci\'{n}ski and L.~Motyka, Phys. Lett. {\bf B438}
(1998) 203.
%
\bibitem{GGSTAR}J. Bartels, A. De Roeck, H. Lotter, Phys. Lett.
{\bf B389} (1996) 742;   J. Bartels , A. De Roeck, C. Ewerz,
H. Lotter, hep-ph/9710500;
A. Bia\l{}as, W. Czy\.z, W. Florkowski, Eur. Phys. J. {\bf C2} (1998) 683;
W. Florkowski, Acta Phys. Polon. {\bf 28} (1997) 2673;
A.~Donnachie, H.G.~Dosch, M. Rueter, Phys.~Rev.~{\bf~D59} (1999) 74011;
M.~Boonekamp et al. hep-ph/9812523.
%
\bibitem{BRODSKY} S.J. Brodsky, F. Hautmann, D.A. Soper, Phys. Rev. {\bf D56} (1997) 6957;
Phys. Rev. Lett. {\bf 78} (1997) 803 (Erratum-ibid. {\bf 79} (1997) 3544);
%
\bibitem{BFKLNL} M. Ciafaloni, G. Camici, Phys. Lett. {\bf B386} (1996) 341;
ibid. {\bf B412} (1997) 396; Erratum -- ibid. {\bf B417}
(1998) 390; hep-ph/9803389; M. Ciafaloni, hep-ph/9709390;
V.S. Fadin, M.I. Kotskii, R. Fiore, Phys. Lett. {\bf B359} (1995) 181;
V.S. Fadin, M.I. Kotskii, L.N. Lipatov, hep-ph/9704267; V.S. Fadin, R. Fiore,
A. Flachi, M. Kotskii, Phys. Lett. {\bf B422} (1998) 287;
V.S. Fadin, L.N. Lipatov, hep-ph/9802290.
%
\bibitem{RESUM} D.A.~Ross, Phys. Lett. {\bf B431} (1998) 161;
G.P.~Salam, JHEP {\bf 9807} (1998), 19; hep-ph/9806482;
M.~Ciafaloni, D.~Colferai hep-ph/9812366; S.J.~Brodsky et al. hep-ph/9901229.
%
\bibitem{KMSG} B. Andersson, G. Gustafson, H. Kharraziha, J. Samuelsson,
Z. Phys. {\bf C71} (1996) 613;   J. Kwieci\'nski, A.D. Martin, P.J. Sutton,
Z. Phys. {\bf C71} (1996) 585.
%
\bibitem{KMSTAS}J. Kwieci\'nski, A.D. Martin, A. Sta\'sto,
Phys. Rev. {\bf D56} (1997) 3991.
%
\bibitem{DLTOTCX} A. Donnachie and P.V. Landshoff, Phys. Lett. {\bf B296}
(1992) 227.
\bibitem{DLDIS}  A. Donnachie and P.V. Landshoff, Z. Phys. {\bf C61} (1994)
139.
%
\bibitem{L3} L3 collaboration, (M. Acciari et al) CERN-EP-98-205 (1998).
%
\bibitem{GGREV}Report of the Working Group on "$\gamma \gamma$ physics",
P. Aurenche and G. A. Schuler (convenerrs), Proceedings of the Workshop
on "Physics at LEP2", editors: G. Altarelli, T. Sj\"ostrand and P. Zwirner,
CERN yellow preprint 96-01.
%
\bibitem{GINZBURG} I.F.Ginzburg, S.L Panfil, V.G. Serbo,
Nucl. Phys. {\bf B296} (1988) 569.
%
%
\bibitem{VMPHOTOP}S. Aid et al., H1 Collaboration,
Nucl. Phys. {\bf B468} (1996) 3; ibid., {\bf B472} (1996) 3;
M. Derrick et al., ZEUS Collaboration, Phys. Lett. {\bf B350} (1996) 120;
J. Breitweg et al., ZEUS Collaboration,
Z. Phys. {\bf C75} (1975) 215.
%

%
\end{thebibliography}
\end{document}